\documentclass[journal]{IEEEtran}

\usepackage{amsmath}
\usepackage{graphicx, epstopdf}
\usepackage{setspace}
\usepackage{acronym}
\usepackage{commath}
\usepackage{mathrsfs}
\usepackage{mathtools}
\usepackage{ifthen}
\usepackage{textcomp}
\usepackage{float}
\usepackage[tight,footnotesize]{subfigure}
\usepackage{cite}
\usepackage{bm}
\usepackage{url}
\usepackage[utf8]{inputenc}
\usepackage{mathtools}
\usepackage{amssymb}
\usepackage{amsthm}
\usepackage{mathrsfs}
\usepackage{cleveref}
\usepackage{bm}
\usepackage{xcolor}
\usepackage{graphicx}

\theoremstyle{definition}

\theoremstyle{definition}

\newcommand{\R}{\mathbb{R}}

\newcommand{\re}{\text{Re}}

\IEEEoverridecommandlockouts

\begin{document}

\title{Probabilistic Stability of Traffic Load Balancing on Wireless Complex Networks}

\author{Giannis Moutsinas\textsuperscript{1}, Weisi Guo\textsuperscript{1,2*}

\thanks{\textsuperscript{1}School of Engineering, University of Warwick, UK. \textsuperscript{1}The Alan Turing Institute, UK. \textsuperscript{*}Corresponding Author: weisi.guo@warwick.ac.uk. This work is funded by EPSRC grant EP/R041725/1.}}


\maketitle

\begin{abstract}
Load balancing between adjacent base stations (BSs) is important for balancing load distributions and improving service provisioning. Whilst load balancing between any given pair of BSs is beneficial, cascade load sharing can cause network level instability that is hard to predict. The relationship between each BS's load balancing dynamics and the network topology is not understood. In this seminal work on stability analysis, we consider a frequency re-use network with no interference, whereby load balancing dynamics doesn't perturb the individual cells' capacity. 

Our novelty is to show an exact analytical and also a probabilistic relationship for stability, relating generalized local load balancing dynamics with generalized network topology, as well as the uncertainty we have in load balancing parameters. We prove that the stability analysis given is valid for any generalized load balancing dynamics and topological cell deployment and we believe this general relationship can inform the joint design of both the load balancing dynamics and the neighbour list of the network. 
\end{abstract}


\section{Introduction}

Load balancing is an important aspect of current and future cellular network operations, homogenizing traffic demand and interference patterns \cite{Bennis17, Chu18, Tall15, Chu18G}. In each base station (BS), load balancing typically involves tuning transmit power and active radio elements to match the traffic demand. When overload with time-sensitive demand, BSs can offload demand to neighbouring BSs, if their demand is relatively low. Load balancing can be implemented between active BSs \cite{Delgado16}, provide support for sleep mode BSs \cite{Guo13JSAC}, user equipments (UEs) in a D2D underlay \cite{Zhang18}, and in wireless sensor networks \cite{Liu18}. Current literature focuses on the delivery mechanism of load balancing and doesn't consider cascade effects across large-scale and hyper-dense networks. We know from other coupled optimisation systems that runaway cascades are possible, see: power control in pairwise coupled BSs \cite{Charalambous12} and in routing \cite{Cholvi10}. In the case of load balancing, this would mean that users are shifted constantly between cells, without a significant improvement in the quality of service, but costing significant spectral inefficiency. Unstable behaviour would be the introduction of new users that cause endless load balancing between cells.

\subsection{Open Challenges}
One challenge with wireless load balancing is that there can be knock-on cascade effects, whereby sharing the load with one neighbouring BS can lead to a run away cascade on the network that is undesirable. For example, it can lead to some BSs that are denying service to many users in order to satisfy load sharing to neighbours. Cascade effects on large scale complex networks (i.e. no. of nodes $N$ is large) that affect stability and equilibrium solutions are difficult to quantify analytically. Recent breakthroughs have shown that there indeed can exist a relationship between local dynamical behaviour and global network structure by compressing the $N$-dimensional dynamics to a $1$-dimensional average behaviour approximation \cite{Gao16}. However, their work examines the average effective behaviour of the whole network and an explicit relationship does not exist universally at the node level. This suffers from covering up discrepancies at the node level. Our own more recent work shows that sequential equilibrium substitution can reveal node level behavioural dynamics \cite{Moutsinas18}, but caveats exist in the application to network topology (e.g. low clustering coefficient).

\subsection{Contribution}
Here, we show for the case of load balancing that an analytical relationship at node level does indeed exist. This is due to the nature of load balancing (e.g. a difference in load demand), which enables direct analytical insight between network topology and the BS load dynamics. We contribute to a rapidly growing literature by deriving an exact analytical relationship for wireless load balancing stability between $N$ coupled BS nodes, relating the behaviour of each BS's load balancing action with the Laplacian of the graph. We also show the conditions for measurement accuracy in the network in order to avoid probabilistic uncertainty causing instability. We believe this general relationship can inform the joint design of both the BS dynamics and the neighbour list for load balancing. \\

\section{System Model}

\subsection{Model Assumptions}
Consider a geographic area covered by $N$ BSs. There are two time scales: long term traffic variations (traffic variation time scale $T$, e.g. seconds), and short term load dynamics under some constant traffic demand (symbol period time scale $t$, e.g. milliseconds). We are primarily concerned with the latter time scale. Each BS $i$ has a load defined by $l_{i}(t) = d_{i}(T)/c_{i}(t)$, the ratio between: (1) the quasi-static long-term traffic demand aggregated across all users $u$ in cell $i$, $d_{i}(T) = \sum_{u} d_{i,u}(T)$; and (2) the BS aggregated area capacity over all users $u$ in cell $i$, $c_{i}(t) = \sum_{u} c_{i,u}(t)$. 

In this seminal paper on stability, we assume that the capacity of each cell is stationary, in that load balancing changes do not dramatically affect the cell capacity (i.e., different from inter-cell inference based load balancing optimisation \cite{Kim12}). This can be justified with frequency re-use patterns and coordinated inter-cell cooperation designed to negate inter-cell interference, both of which are actively researched and utilized technologies in hyper-dense scenarios \cite{Agiwal16}. We do not consider user-level experience in this initial paper, and rather focus on network level stability.

\begin{figure}[t]
    \centering
    \includegraphics[width=1.00\linewidth]{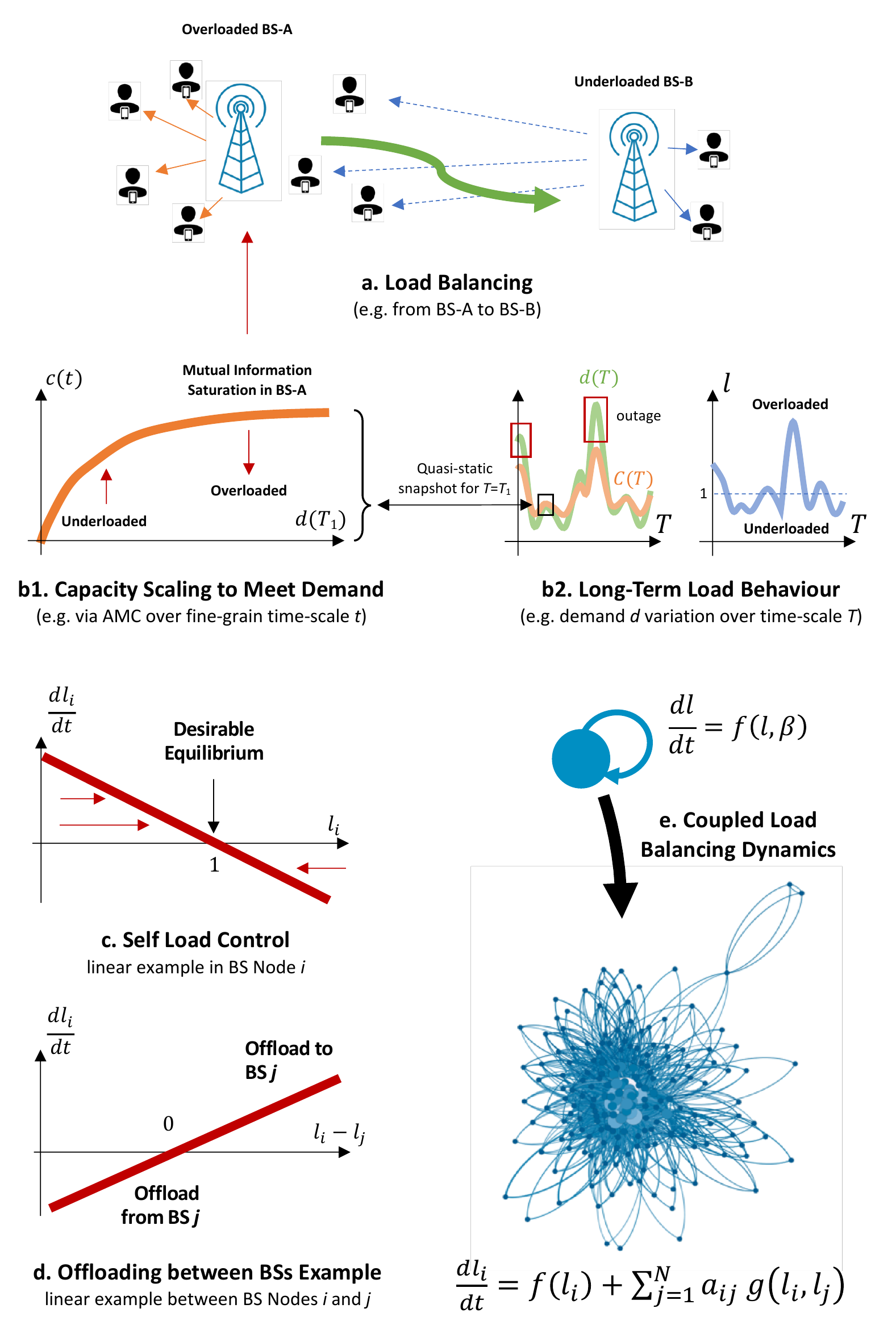}
    \caption{{\bf Wireless Network Load Balancing Dynamics: a) illustration of load balancing between frequency-reuse cells, b) capacity saturates for high loads drives load balancing requirement, c) an example of the load dynamic control for a single cell, d) an example of the load balancing action between two cells, and e) a complex network of load balancing between $N$ cells. }}
    \label{fig:1}
\end{figure}

We are interested in the transient dynamics and stability of load balancing at time scale $t$, for a particular demand $d_{i}(T=T_{1})$ - see Figure~\ref{fig:1}b. As such, we do not yet examine the user-level aspects of demand change, scheduling and propagation dynamics, nor user flow \cite{Liakopoulos18}. Suffice to say, we acknowledge that the wireless capacity depends on user location and PHY/MAC protocols, but for this letters, we simply model each BS as being under a certain random demand value and able to deliver a certain capacity profile.

\subsection{Linear Example of Load Balancing Dynamics}
Within the quasi-static traffic demand regime, the BS capacity $c_{i}(t)$ reacts to the demand $d(T=T_{1})$ using adaptive modulation and coding (AMC) - see Figure~\ref{fig:1}b1. However, the mutual information of discrete modulation constellations will saturate \cite{Guo13WCNC}, and therefore as the load exceeds 1, load balancing is necessary in order to avoid outage (see Figure~\ref{fig:1}b2). As a demonstration example, we assume an ideal and simple linear load scaling between capacity and demand\footnote{In Section III, we show that our results hold for any general dynamics.}. As such, the load dynamics in cell $i$ can be described by (see Figure~\ref{fig:1}c):
\begin{align}
 \dot{l}_i = f(l_{i}) = \beta(1-l_{i}),
 \label{eq_telecom_self}
\end{align} where a desirable equilibrium for \textit{maximum service efficiency} is at $l_{i}=1$ (fully loaded). The parameter $\beta$ accounts for the efficiency of the scaling process.

Each BS may have a list of adjacent BSs neighbours that it can share load with, and we can think of this virtual coupling of loads as through the $a_{ji}$ connectivity matrix (e.g. a BS load sharing neighbour list). The dynamics of the offloading process can be described by the difference in the BSs' loads (see Figure~\ref{fig:1}d):
\begin{align}
 \dot{l}_i = g(l_{i},l_{j}) = \gamma(l_{j} - l_{i}),
 \label{eq_telecom_couple}
\end{align} with an offloading rate $\gamma$. 

As such, the overall dynamics in terms of wireless traffic load can be simply described by (see Figure~\ref{fig:1}e):
\begin{align}
 \dot{l}_i = \beta(1-l_{i}) + \sum_{j=1}^{N}a_{ji} \gamma(l_{j} - l_{i}),
 \label{eq_telecom_load}
\end{align} where $a_{ji}=(A)_{ji}$. Here, we note that the dynamics due to cascades is $N$-dimensional, which makes direct prediction on stability challenging when $N$ is large. \\

\section{Stability Analysis of General Dynamics}

Setting aside the linear dynamics example in Section II, here we look at the general case
\begin{equation}
 \dot l_i = f(l_i) + \sum_{j=1}^N a_{ji} g(l_j-l_i), 
 \label{eq_Dyn_Sys}
\end{equation} where $f,g$ are twice differentiable functions and $g(0)=0$. In order to understand stability, we need to look at the linearization of the system. As such, we write $g(x) = \gamma x + O(x^2)$ as the Taylor expansion, where $O(\cdot)$ is the big O notation which bounds asymptotic behaviour at $x=0$.

We denote $ L = (l_1,\dots,l_N)$ and we write equation \eqref{eq_Dyn_Sys} as
\begin{equation}
 \dot {L} = F(L).
 \label{eq_Dyn_Sys_one_line}
\end{equation}
Let $\mathbf 1 = (1,\dots,1)$, then it is straightforward to check that if $r$ is a root of $f$, i.e. $f(r)=0$; then $r\mathbf 1$ is an equilibrium of the dynamical system.

For any load balancing dynamics stated in Eq.\eqref{eq_Dyn_Sys}, we know that there exists an equilibrium solution at $r\mathbf 1$. For linear dynamics (see Section II), this is the \textbf{only} equilibrium solution. As such, we provide the analysis for stability around this equilibrium solution. 

In order to determine the stability of the equilibrium we compute the eigenvalues of the Jacobian at the equilibrium. Let $F_i$ be the $i$-th component of the function $F$ of equation
\eqref{eq_Dyn_Sys}, then we have 
\begin{equation}\begin{split}
\left.\frac{\partial}{\partial l_i} F_i (L) \right|_{L=r\mathbf{1}} &= f'(r) - \sum_{j=1}^N a_{ji} g'(0) \\
 &= f'(r) - \gamma w_i.
 \label{eq_slope}
\end{split}\end{equation} where we have defined $w_i = \sum_{j=1}^N a_{ji}$ and by the assumptions on $g$ it holds that $g'(0)=\gamma$.

When $k\ne i$ we have
$$ \left.\frac{\partial}{\partial l_k} F_i (L) \right|_{L=r\mathbf{1}} = \sum_{j=1}^N \delta_{jk} a_{ji} g'(0) = \gamma a_{ki}, $$
where $\delta_{ki}$ is the Kronecker delta. This equation together with equation \eqref{eq_slope} shows that the Jacobian has the form
\begin{equation}
 J(r \mathbf 1) = f'(r) \text{Id} - \gamma D + \gamma A^T = f'(r) \text{Id} - \gamma \Lambda^T,
 \label{eq_stability}
\end{equation} where $\text{Id}$ is the identity matrix, $D$ is the weighted in-degree matrix and $\Lambda$ the weighted in-Laplacian of the graph and $\Lambda^T$ its transpose. Notice that the spectrum of $J(r \mathbf 1)$ is a spectral shift of the spectrum of $\gamma \Lambda^T$. Remember that $\Lambda$ and $\Lambda^T$ have the same spectrum.

\subsection{Gershgorin Circle Theorem}
For the Laplacian it is known that $0$ is an eigenvalue and that all eigenvalues have non-negative real part. The first assertion is a direct implication of the relation $\Lambda^T\cdot\mathbf{1}=0$. The second assertion a consequence of Gershgorin circle theorem, \cite{varga2011gersgorin}. For each row of the matrix we construct the disc that has the diagonal element as centre and the sum of the absolute values of the remaining elements as radius, we call each of these discs a \textit{Gershgorin disc}. Gershgorin's theorem states that each Gershgorin disc contains at least one eigenvalue of the matrix.

Because a matrix and its transpose have the same eigenvalues, we can do the same with the columns instead of the rows. In the case of the Laplacian matrix, since the sum of a row is zero and the diagonal elements are all non-negative, each disc has centre on the positive real axis and is tangent to the imaginary axis.

\subsection{Stability Scenarios for Various Dynamics}
Let $\mu_i$ denote the eigenvalues of $ J(r \mathbf 1)$ and $\lambda_i$ denote the eigenvalues of $\Lambda$, the relation between them is $\mu_i = f'(r) - \gamma \lambda_i$. The equilibrium $r \mathbf{1}$ is stable if $\re(\mu_i)>0$ for all $i$. Then from the discussion on the eigenvalues of $\Lambda$ we deduce the following:
\begin{itemize}

    \item \textbf{Default Load Balancing:} If $f'(r) < 0$ and $\gamma \ge 0$, then the equilibrium $r\mathbf 1$ is asymptotically \textit{stable}. This scenario is the default load balancing setup. As such, in this default case, the dynamics (e.g. $f(\cdot), \gamma$) only affect how resilient the stable system is to faults and how fast it contracts, but not the stability itself.
    
    Since the system \eqref{eq_telecom_load} is linear and we know that the largest eigenvalue of the Jacobian is $-\beta$, we know that regardless the initial condition the system contracts to the equilibrium point as $e^{-\beta t}$.

    \item If $f'(r) < 0$ and $\gamma < 0$, then the equilibrium $r\mathbf 1$ is asymptotically \textit{stable} if $|f'(r)| > |\gamma| \,\rho$ and asymptotically unstable if $|f'(r)| < |\gamma| \,\rho$, where $\rho = \max\{ \re(\lambda_i) \}$. This scenario is for when load must be given to more heavily loaded BSs, as a potential first step before considering \textit{sleep mode} \cite{Guo13JSAC}.

    \item If $f'(r) = 0$ and $\gamma < 0$, then the equilibrium $r\mathbf 1$ is asymptotically \textit{unstable}. This scenario is similar to the above \textit{sleep mode} case.
    
    \item If $f'(r) = 0$ and $\gamma \ge 0$, then we \textit{cannot determine the stability} of the equilibrium $r\mathbf 1$ just by looking at the eigenvalues of the Jacobian. This scenario might be suitable for \textit{multi-hop routing}.
    
    \item If $f'(r) > 0$, then the equilibrium $r\mathbf 1$ is asymptotically \textit{unstable}. This scenario is not applicable to most telecommunication dynamics.
\end{itemize}
For the purpose of load balancing as described in this paper, we are only interested in the case of $f'(r) < 0$ and $\gamma > 0$, which when referring to the system considered in Eq.\eqref{eq_telecom_load}, it implies that the equilibrium is always asymptotically stable. \\

\section{Probabilistic Uncertainty}

In realistic networks, each BS can have different load balancing mechanisms. Measurement of the parameters and load flow are subject to measurement noise and can affect stability. We investigate this by adding a random variable to the load scaling gradient $\beta$ and load balancing rate $\gamma$ (see Eq.\eqref{eq_telecom_load}). As such, the dynamical system is $\dot{l}_i = (\beta + \zeta_i)(1-l_i) + \sum_{j=1}^N a_{ji} (\gamma + \xi_{ji})(l_j-l_i)$, where $\zeta_i$ and $\xi_{ji}$ are random variables of known distribution. In this case we have $f'(r) = -\beta$. The entries of the Jacobian matrix are
\begin{align*}
    (J)_{ii} &= - \beta - \zeta_i - \sum_{j=1}^N a_{ji} (\gamma + \xi_{ji})\\
    (J)_{ij} &= a_{ji} \gamma + a_{ji} \xi_{ji}.
\end{align*}
We define $\theta_i = \sum_{j=1}^N a_{ji} \xi_{ji} $, then the Jacobian becomes 
$$ J = -\beta \, \text{Id} - \gamma\, \Lambda^T - Z - \Theta - \Xi, $$
with $Z = \text{diag}(\zeta_1,\dots,\zeta_N)$, $\Theta=\text{diag}(\theta_1,\dots,\theta_N)$ and $(\Xi)_{ij} = a_{ji} \xi_{ji}$.

Just like before, the system is stable if and only if all the eigenvalues of $J$ have negative real part. We use the Gershgorin circle theorem to get a bound on that probability. Let 
\begin{equation*}
    s_i = -\beta - \zeta_i + \sum_{j=1}^N a_{ji} (|\gamma + \xi_{ji}| - \gamma - \xi_{ji} ),
\end{equation*}
then if for all $i$, $s_i<0$ then the system is stable.

In this case, the \textit{stability is probabilistic},  i.e., the probability that the system is stable is bounded from below by $\prod_{i=1}^N \mathbb{P}(s_i>0)$. The exact form of this bound will depend on the distribution of the random variables $\zeta_i$ and $\xi_{ji}$. \\

As an example we will look at the case of Default Load Balancing where the we know the values of $\beta$ and $\gamma$ up to a uniform measurement error.
Let $\beta>0$, $\gamma>0$, $\zeta_i \sim \text{Uniform}[-b,b]$ and  $\xi_{ji} \sim \text{Uniform}[-c,c]$, with $c>\gamma$. Let $X_i = |\gamma + \xi_{ji}| - \gamma - \xi_{ji} $ Since $\mathbb{P}(\gamma + \xi_{ji} < 0) = \tfrac{1}{2}(1-\gamma/c) $, the random variables $X_i$ take the values
\begin{equation*}
    X_i \sim \begin{cases} 
      0, & \text{with probability }\tfrac{1}{2}(1+\gamma/c) \\
      \text{Uniform}[0,2(c-\gamma)], & \text{with probability }\tfrac{1}{2}(1-\gamma/c).
   \end{cases}
\end{equation*}
We denote $Y=\sum_{i=1}^N X_i$. The random variable $Y$ is the sum of $n$ uniform random variables with probability $\binom{N}{n}(\tfrac{1}{2}(1-\gamma/c))^n(\tfrac{1}{2}(1+\gamma/c))^{N-n}$.
The sum of independent uniform random variables follows the Irwin-Hall distribution. This implies that the PDF of $Y$ is
\begin{align*}
    f_Y(x) &= \frac{1}{2^N}\sum_{n=1}^N \binom{N}{n}(1-\gamma/c)^n(1+\gamma/c)^{N-n} \\ &\phantom{=}\times \frac{1}{(n-1)!}\!\!\sum_{k=0}^{\left\lfloor \frac{x}{2(c-\gamma)} \right\rfloor}(-1)^k \binom{n}{k}\left(\frac{x}{2(c-\gamma)}-k\right)^{n-1}.
\end{align*}

We can now write $s_i = -\beta - \zeta_i + Y_i$ and ask what is the probability that $s_i$ is positive. The random variable $-\beta - \zeta_i$ is uniform on $[-\beta-b,-\beta+b]$ and PDF of the sum of two random variables is the convolution of the PDFs, we get
\begin{align*}
    \mathbb{P}(s_i<0) = \frac{1}{2b} \int_{-\infty}^0\int_{x+\beta-b}^{x+\beta+b} f_Y(s)\, ds\, dx.
\end{align*}
This integral cannot be written in a simple form, but it can easily be evaluated numerically. Finally, the probability that every $s_i$ is negative is $(\mathbb{P}(s_i<0))^n$, which bounds from below the probability that the system is stable.

Note that we looked at the case where $c>\gamma$, which corresponds to a measurement error that is of the same order of magnitude as the measurement itself. If we assume that the error is smaller, i.e. $c\le\gamma$ then the $\gamma + \xi_{ji}$ cannot be negative, which implies that the system is always stable. \\

\begin{figure}[t]
    \centering
    \includegraphics[width=1\linewidth]{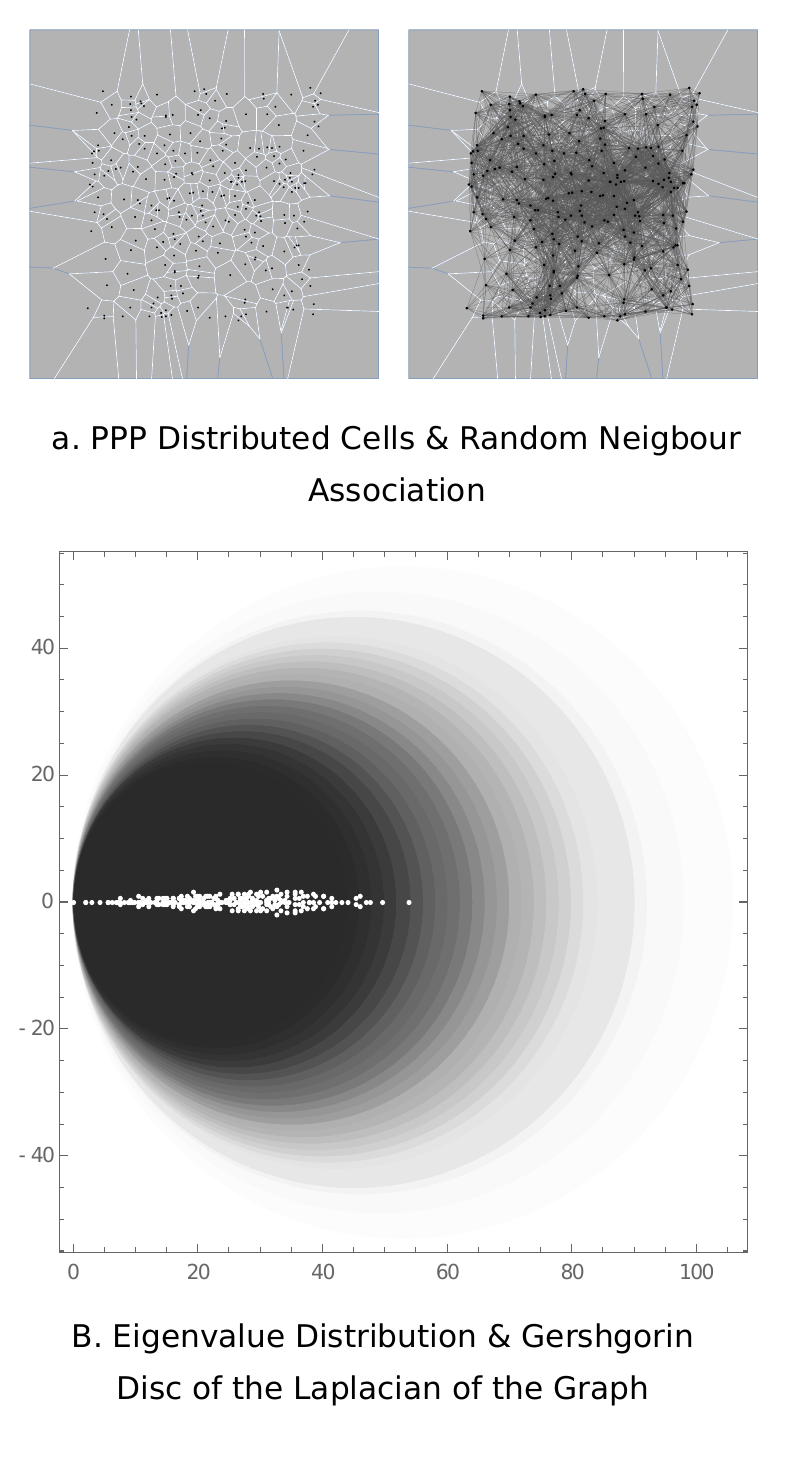}
    \caption{{\bf Distribution of Eigenvalues for PPP Distributed Network.}}
    \label{fig:2}
\end{figure}
\begin{figure}[t]
    \centering
    \includegraphics[width=0.99\linewidth]{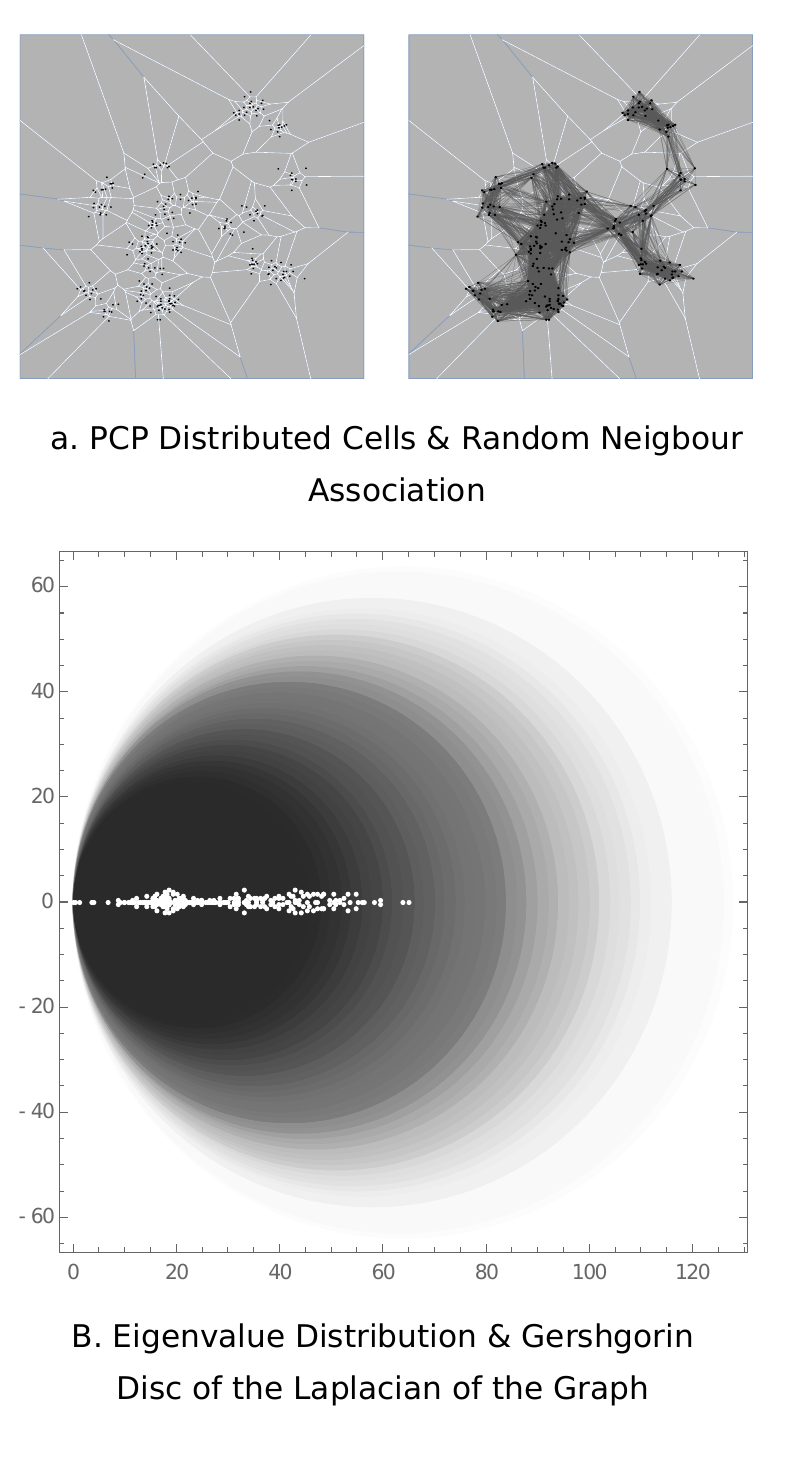}
    \caption{{\bf Distribution of Eigenvalues for PCP Distributed Network.}}
    \label{fig:3}
\end{figure}

\section{Capacity Stability Analysis}

A smooth invertible transformation of a dynamical system does not change the stability of its equilibria. In particular let $\phi_i:\R\to\R$ be invertible, twice continuously differentiable functions and let us define $c_i=\phi_i(l_i)$. The dynamical system for $c_i$'s is given by the equations using chain rule $\dot c_i = \phi_i'\big(l_i\big)\dot l_i$:
\begin{equation}\begin{split}
 \dot c_i = &\phi_i'\big(\phi_i^{-1}(c_i)\big) \times \\
 &\left( f\big(\phi_i^{-1}(c_i)\big) + \sum_{j=1}^N a_{ji} g\big(\phi_j^{-1}(c_j) - \phi_i^{-1}(c_i)\big)\right).
 \label{eq_Dyn_Sys_Cs}
\end{split}\end{equation}
From the previous discussion we know that the point $(\phi_1(r),\dots,\phi_N(r))$ is an equilibrium of the system \eqref{eq_Dyn_Sys_Cs} that corresponds to the equilibrium $r\mathbf 1$ of the system \eqref{eq_Dyn_Sys}. The stability of this equilibrium is the same as the stability of $r\mathbf{1}$.

Now we can apply the previous analysis in the case of of load-balancing of BSs, governed by the dynamical system \eqref{eq_telecom_load}. Notice that this system is linear, which implies that there is only one equilibrium and by the previous analysis we know that when $\gamma,\;\beta>0$, this equilibrium is asymptotically stable.

We assume that $\phi_i(l_i)=d_i/l_i$. This implies that $\phi_i^{-1}(c_i)=d_i/c_i$ and $\phi_i'(l_i)=-d_i/l_i^2$. Then the system \eqref{eq_Dyn_Sys_Cs} becomes
\begin{align}
    \dot c_i &= -\frac{c_i^2}{d_i}\left( \beta \left( 1-\frac{d_i}{c_i} \right) + \sum_{j=1}^N a_{ji} \gamma \left( \frac{d_j}{c_j} - \frac{d_i}{c_i} \right) \right) \nonumber \\
    &=  \beta c_i \left( 1-  \frac{c_i}{d_i} \right) + \sum_{j=1}^N \gamma \, a_{ji} c_i  \left(1 - \frac{c_i d_j}{c_j d_i} \right).
    \label{eq_Dyn_Sys_BSs}
\end{align}
At first glance it seems that the above equation implies that the self-dynamics of a BS is given by $f(c_i)=\beta c_i(1-c_i/d_i)$ and it has two equilibria, $d_i$ which is stable and $0$ which is unstable. The equilibrium $(d_i,\dots,d_N)$ corresponds to the stable equilibrium of the system \eqref{eq_telecom_load} and from this we deduce that it is not just asymptotically stable but also a global attractor of the system. Moreover, we get that it contracts with the same speed, i.e. $e^{-\lambda t}$. The equilibrium $0$, however, is not an admittable one because it appears also as a denominator and in this case the right-hand side of \eqref{eq_Dyn_Sys_BSs} cannot be evaluated. In a sense the $0$ ``equilibrium'' of the system \eqref{eq_Dyn_Sys_BSs} corresponds to infinity in the system  \eqref{eq_telecom_load}.  \\

\section{Results}

We present results for differing Poisson Point Process (PPP) and Poisson Cluster Process (PCP) generated random complex networks \cite{Dhillon18}, where nodes are omni-directional BS sites and links are offloading relations. We connect the nodes in accordance to a random network, whereby a percolation control parameter $R$ and a probability of connecting $P$ is used to determine if adjacent BSs can offload to each other. Traffic and capacity values are not needed because we know from earlier that the system is always stable and the load dynamics only affect the rate of contraction and resilience to faults, but not its asymptotic stability. This is an important insight. 

Fig.\ref{fig:2}a show the PPP Voronoi plots along with BS load balancing network. The results in Fig.\ref{fig:2}b demonstrate that, as we expected, the eigenvalues of the Laplacian are all in the positive real half-plane. Therefore the whole load balancing network is always stable in this case. Fig.\ref{fig:3}a show the PCP Voronoi plots along with BS load balancing network. The results in Fig.\ref{fig:3}b demonstrate that, as we expected, the eigenvalues of the Laplacian are all in the positive real half-plane. Therefore the whole load balancing network is always stable in this case. \\

\section{Conclusion \& Future Work}

In this paper, we show the stability criteria that links the generalized load balancing dynamics ($f(\cdot),\gamma$) with the maximum eigenvalue of the weighted in-Laplacian of the adjacency matrix ($\rho$). We prove that default load balancing networks are always asymptotically stable, irrespective network topology and the balancing dynamics (linear or otherwise). However, we observe that for other forms of balancing actions, the stability is not ensured. We also present the probabilistic stability in the face of heterogeneous uncertainty among the load balancing actions. We showed that given uncertainty in the load balancing actions, as long as the system measurement accuracy is better the underlying noise process, the system is stable.

We believe this general relationship can inform the joint design of both the base station (BS) dynamics and the BS interaction network. Whilst this seminal work on stability analysis considered a frequency re-use network with no interference, future work will consider the effects of interference, sleep mode, user entry/exit demand dynamics \cite{Liakopoulos18}, and their influence on stability.

\bibliographystyle{IEEEtran}
\bibliography{IEEEabrv,CL_Ref}

\end{document}